\documentclass[aps,pra,twocolumn,amsmath,showpacs,superscriptaddress]{revtex4}

\usepackage{graphicx}
\usepackage{amssymb}
\usepackage{pifont}
\usepackage{color}
\usepackage[mathscr]{euscript}
\usepackage{mathrsfs}

\DeclareMathAlphabet{\mathpzc}{OT1}{pzc}{m}{it}
\def\vol{\mathpzc{V}}

\begin{document}

\title{Bose-Einstein condensation in dark power-law laser traps}

\author{A. Jaouadi}
\affiliation{Universit\'e Paris-Sud, Institut des Sciences Mol\'{e}culaires d'Orsay (ISMO), F-91405 Orsay, France.}
\affiliation{CNRS, Orsay, F-91405 France.}
\affiliation{Laboratoire de Spectroscopie Atomique, Mol\'{e}culaire et Applications (LSAMA), Department of Physics, Faculty of Science of Tunis, University of Tunis El Manar, T-2092 Tunis, Tunisia.}

\author{N. Gaaloul}
\affiliation{Institut f\"ur Quantenoptik, Welfengarten 1, Gottfried Wilhelm Leibniz Universit\"at, D-30167 Hannover, Germany.}

\author{B. Viaris de Lesegno}
\affiliation{CNRS, Laboratoire Aim\'{e} Cotton (LAC), F-91405 Orsay, France.}
\affiliation{Universit\'e Paris-Sud, Orsay, F-91405 France.}

\author{M. Telmini}
\affiliation{Laboratoire de Spectroscopie Atomique, Mol\'{e}culaire et Applications (LSAMA), Department of Physics, Faculty of Science of Tunis, University of Tunis El Manar, T-2092 Tunis, Tunisia.}

\author{L. Pruvost}
\affiliation{CNRS, Laboratoire Aim\'{e} Cotton (LAC), F-91405 Orsay, France.}
\affiliation{Universit\'e Paris-Sud, Orsay, F-91405 France.}

\author{E. Charron}
\affiliation{Universit\'e Paris-Sud, Institut des Sciences Mol\'{e}culaires d'Orsay (ISMO), F-91405 Orsay, France.}
\affiliation{CNRS, Orsay, F-91405 France.}

\date{\today}

\begin{abstract}
We investigate theoretically an original route to achieve Bose-Einstein condensation using dark power-law laser traps. We propose to create such traps with two crossing blue-detuned Laguerre-Gaussian optical beams. Controlling their azimuthal order $\ell$ allows for the exploration of a multitude of power-law trapping situations in one, two and three dimensions, ranging from the usual harmonic trap to an almost square-well potential, in which a quasi-homogeneous Bose gas can be formed. The usual cigar-shaped and disk-shaped Bose-Einstein condensates obtained in a 1D or 2D harmonic trap take the generic form of a ``finger'' or of a ``hockey puck'' in such Laguerre-Gaussian traps. In addition, for a fixed atom number, higher transition temperatures are obtained in such configurations when compared with a harmonic trap of same volume. This effect, which results in a substantial acceleration of the condensation dynamics, requires a better but still reasonable focusing of the Laguerre-Gaussian beams.
\end{abstract}
\maketitle

\section{Introduction}

Quantum information processing and matter-wave interferometry are two applications of atom optics which demand an unprecedented level of quantum control of the external and internal degrees of freedom of atoms and molecules. They require the implementation of miniaturized atomic and molecular traps in order to minimize the number of accessible translational energy levels. These traps can be realized using either an inhomogeneous magnetic field created by micro-fabricated circuits on an atom chip\,\cite{Fortagh2007} or the inhomogeneous electric field of a laser beam profile\,\cite{Miller1993}. The first technique, which allows for the implementation of complex trap configurations of almost arbitrary geometry\,\cite{Charron2006}, is limited to specific Zeeman sublevels and may suffer from decoherence due to the proximity with the chip surface. With the second technique, any kind of internal atomic or molecular state may be trapped in a non-dissipative way, but this approach usually suffers from a lack of flexibility since the shape of the trapping potential is not easily reconfigurable\,\cite{Gaaloul2006}.

However, the use of programmable diffractive optical elements has recently brought a new flexibility for the optical manipulation of single atoms\,\cite{Bergamini2004} and of cold atomic ensembles\,\cite{Boyer2006-Viaris2007}.

With atomic traps realized using the optical dipole force, the atoms can be attracted or repelled from high intensity regions, depending on the sign of the laser detuning with respect to the atomic transition frequency. With blue-detuned traps, where the light frequency is tuned above the atomic resonance, it is possible to avoid many of the possible sources of decoherence since the atoms are confined in regions of minimum light intensity. In this case, spontaneous photon scattering events are relatively rare, providing low heating rates. The light-induced shifts of the atomic energy levels are also reduced, and low laser powers are usually sufficient to trap a large collection of atoms during a relatively long time\,\cite{Fatemi2008}.

Dark hollow laser beams offer interesting perspectives in this context due to the development of efficient approaches for generating these types of optical configurations with programmable\,\cite{Rhodes2006,Fatemi2006-Olson2007} and dynamically reconfigurable\,\cite{Franke-Arnold2007} holograms.

\begin{figure}[!t]
\centering
\includegraphics*[width=8.6cm,clip=true]{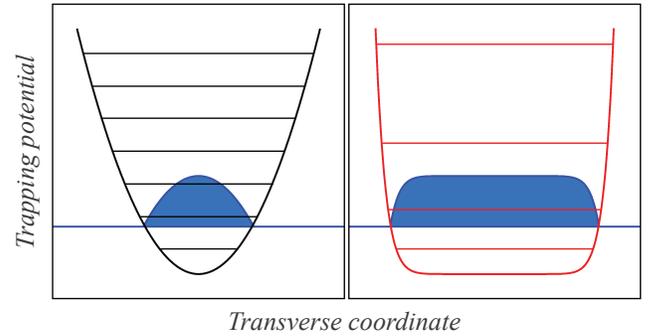}
\caption{(Color online) Schematic cut of the trapping potential $V_{\ell}(\rho)$ from Eq.\,(\ref{Eq:Dipole}) along the $x$ axis for $\ell=1$ (harmonic case, left) and for $\ell=6$ (right) with their associated energy level structures. The blue shapes represent the expected Thomas-Fermi condensate atomic densities.}
\label{fig:schematic}
\end{figure}

Laguerre-Gaussian (LG$_{p}^{\ell}$) optical beams\,\cite{Clifford1998} are such circularly symmetric hollow laser modes. They are characterized by a radial index $p$ and an azimuthal index $\ell$\,\cite{NoteLG}. These beams carry a well defined orbital angular momentum $\ell\hbar$ along their propagation axis. This property was used recently for the coherent preparation of atomic vortex states in a Bose-Einstein condensate (BEC) of sodium atoms\,\cite{Phillips2006} and to induce the persistent flow of a BEC confined in a toroidal trap\,\cite{Phillips2007}.

Here, we report on the transition temperature and on the growth dynamics of trapped interacting dilute Bose gases in blue-detuned Laguerre-Gaussian laser beams. In the dark region surrounding the beam propagation axis, the atoms are confined in a quasi power-law trapping potential $V_{\ell}(\rho) \varpropto \rho^{2\ell}$, where $\rho=\sqrt{x^2+y^2}$ is the radial distance to the beam center\,\cite{Arlt2000}. Typical shapes of such trapping potentials are shown in Fig.\,\ref{fig:schematic} in the case of a harmonic trap $(\ell=1)$, and for $\ell=6$, along with the Thomas-Fermi atomic densities expected for large condensate occupation numbers. These traps are characterized by very different energy level structures which will necessarily affect the condensation process. In addition, for large condensate occupation numbers, the atomic density should reflect the shape of the potential, yielding very different distributions for different values of $\ell$, as seen schematically in Fig.\,\ref{fig:schematic}. For $\ell \rightarrow \infty$, a quasi square-well potential is obtained, providing a homogeneous trap. In practice, for $\ell \geqslant 4$ the condensate is characterized by an almost constant density over the entire trap volume.

\section{Description of the trap}

\begin{figure}[!t]
\centering
\includegraphics*[width=8.6cm,clip=true]{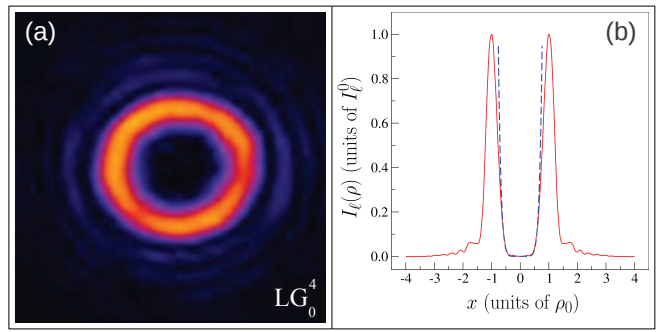}
\caption{(Color online) (a) Light intensity of a LG$_0^4$ mode (with $\ell=4$) as recorded by a CCD camera close to the focal point (see\,\cite{Mestre2010} for details). (b) Red solid line: associated averaged intensity profile. Blue dashed line: simple power-law variation proportional to $\rho^{8}$.}
\label{fig:LG04mode}
\end{figure}

In order to create a power-law laser trap we propose to use Laguerre-Gaussian beams, more precisely LG$_{0}^{\ell}$ modes with $p=0$. These modes are circularly symmetric, and their intensity profile can be written as
\begin{equation}
I_{\ell}(\rho) = \frac{2}{\pi\ell!}\,\frac{P}{w_0^2}\,\left(\frac{2\rho^2}{w_0^2}\right)^{\ell}\exp\left(-\frac{2\rho^2}{w_0^2}\right)\,,
\label{Eq:Intensity-exact}
\end{equation}
where $\ell$ is the order of the mode, $P$ the beam power and $w_0$ the beam waist. Inside the Rayleigh range $|z| \ll z_R$ with $z_R = \pi w_0^2/\lambda$, the beam waist can be considered as constant. $\lambda$ denotes here the radiation wavelength. As shown in Fig.\,\ref{fig:LG04mode}, the laser profile consists in a ring of light with the maximum intensity
\begin{equation}
I_{\ell}^0 = \frac{2\ell^{\ell}}{\pi\ell!e^{\ell}}\,\frac{P}{w_0^2}\,,
\label{Eq:Intensity-max}
\end{equation}
located at $\rho = \rho_0 = w_0\sqrt{\ell/2}$.

Many methods are available for creating Laguerre-Gaussian modes. One of them consists in applying a helical phase $\chi(\phi)=\ell\,\phi$ to the wavefront of a Gaussian beam. This transforms the Gaussian beam into a nearly-pure Laguerre-Gaussian mode. Experimentally, it is performed with a phase-hologram which can be realized for example with a liquid crystal spatial light modulator. Details about the method and the generated modes are given in Ref.\,\cite{Rhodes2006} and in Ref.\,\cite{Mestre2010}. As can be seen in Fig.\,\ref{fig:LG04mode}(b), the averaged measured intensity profile follows a clean power-law variation $I_\ell(\rho) \propto \rho^{2\ell}$ inside the ring of light. In addition, near the ring center the light intensity, which should approach zero according to Eq.\,(\ref{Eq:Intensity-exact}), quickly reaches the background noise of the CCD camera.

Far- and blue-detuned from the atomic resonance, the center of the intensity profile of Eq.\,(\ref{Eq:Intensity-exact}) can be used to trap neutral atoms. The associated dipole optical potential can be written as
\begin{equation}
V_\ell(\rho) = \frac{\hbar\Gamma_{\!\!s}^2}{8\delta}\,\frac{I_\ell(\rho)}{I_s}
\label{Eq:Dipole-exact}
\end{equation}
in the framework of the usual two-level approximation\,\cite{Miller1993}. $\Gamma_{\!\!s}$ is the spontaneous atomic emission rate, $\delta$ the laser detuning, and $I_s$ the saturation intensity\,\cite{NoteRb}. This trapping potential is therefore characterized by a potential barrier of height
\begin{equation}
V_\ell^0 = \frac{\hbar\Gamma_{\!\!s}^2}{8\delta}\,\frac{I_\ell^0}{I_s}
\label{Eq:Dipole-max}
\end{equation}
located at $\rho = \rho_0$. In addition, well inside the waist radius $(\rho \ll w_0)$, and for atoms having a temperature $k_BT \ll V_\ell^0$, this potential is, in a good approximation, given by the expression
\begin{equation}
V_\ell(\rho) \simeq \frac{2^{\ell}}{4\pi\ell!}\,\frac{\,\hbar\Gamma_{\!\!s}^2}{\delta I_s}\,\frac{P}{w_0^2}\;\left(\frac{\rho}{w_0}\right)^{2\ell}\,,
\label{Eq:Dipole}
\end{equation}
which follows a simple \textit{even} power-law variation.

In order to create a 3-dimensional trap, we have considered an all-optical configuration consisting of two perpendicularly crossing LG$_0^\ell$ laser modes. The polarizations of these two beams are chosen to be orthogonal to avoid any interference pattern. The first beam, circularly symmetric, propagates along the $z$ direction and provides trapping in the ($x,y$) plane. Trapping in the third dimension is provided by another strongly elongated Laguerre-Gaussian beam shaped elliptically in the form of a light sheet and propagating in the $x$ direction. With this beam configuration, the corresponding potential near the trap center is described by
\begin{equation}
V(\rho,z) = U_{\!\bot}\,\rho^{\alpha} + U_{\!z}\,z^{\beta}\,,
\label{Eq:Potential}
\end{equation}
where $\alpha$ and $\beta$ are two \textit{even} integers. As seen in Eq.(\ref{Eq:Dipole}), $U_{\!\bot}$ and $U_{\!z}$ are simple functions of the laser parameters $P$, $\delta$ and $w_0$ of the two beams.

\section{Bose-Einstein condensation in a power-law trap}

\subsection{Geometrical aspects}

In the mean-field regime, a BEC formed in such a trap can be described by the  normalized macroscopic condensate wave function $\Psi_c(\rho,z)$ solution of the Gross-Pitaevskii equation\,\cite{Pitaevskii1961-Gross1961-Gross1963}
\begin{equation}
-\frac{\hbar^2}{2m}\nabla^2\Psi_c+V(\rho,z)\Psi_c+g\,N_c\left|\Psi_c\right|^2\Psi_c=\mu_c\Psi_c\,,
\label{Eq:GPE}
\end{equation}
where $m$ denotes the atomic mass and $\mu_c$ is the chemical potential related to the condensate number $N_c$. The non-linear term $g\,N_c\left|\Psi_c\right|^2$ describes the mean field two-body interaction whose strength
\begin{equation}
g=\frac{4\pi\hbar^2a_s}{m}
\label{Eq:interaction}
\end{equation}
is proportional to the $s$-wave scattering length $a_s$\,\cite{Kempen2002}.

One of the main parameters which controls the condensation process is the trap confinement. Indeed, for any trap, the condensation process is taking place when the atomic density is approximately one particle per cubic thermal de Broglie wavelength. More precisely, it can be shown\,\cite{Pethick2008} that the \textit{peak} atomic density $n(0)$ in the center of the trap and at the onset of condensation is
\begin{equation}
n(0)=\frac{\zeta\left(\frac{3}{2}\right)}{\lambda_T^3}\,,
\label{Eq:n0}
\end{equation}
where $\zeta(x)$ denotes the Riemann zeta function\,\cite{Abramowitz1964} and
\begin{equation}
\lambda_T=\frac{2\pi\hbar^2}{\sqrt{mk_BT}}
\label{Eq:lambdaT}
\end{equation}
is the thermal de Broglie wavelength at temperature $T$. The trap confinement, having a strong influence on $n(0)$, has consequently a large impact on the transition temperature.

In order to study the effects of the shape of the potential, \textit{and of this shape only}, on the condensation process, we have decided to fix the average atomic density, and therefore to fix the volume of space occupied by the condensate. For this purpose, we first define the volume of the trap where $V(\rho,z) \leqslant \varepsilon$, $\varepsilon$ being an arbitrary energy. This volume can be written as
\begin{equation}
\vol_{T}(\varepsilon) \;=\; \int_{V(\rho,z) \leqslant \varepsilon}\!\!\!\!\!\!\!\!d^3r \;=\; \frac{2\pi\,C_{\alpha\beta}}{\Gamma\left(\eta+\frac{1}{2}\right)}\,\varepsilon^{\eta-\frac{1}{2}}\,,
\label{Eq:pseudo-volume}
\end{equation}
where
\begin{equation}
C_{\alpha\beta} = U_{\!\bot}^{-\frac{2}{\alpha}}\,U_{\!z}^{-\frac{1}{\beta}}\;\Gamma\left(\frac{2}{\alpha}+1\right)\,\Gamma\left(\frac{1}{\beta}+1\right)\,,
\label{Eq:Cab}
\end{equation}
and where
\begin{equation}
\eta = \frac{2}{\alpha} + \frac{1}{\beta} + \frac{1}{2}
\label{Eq:eta}
\end{equation}
denotes an important parameter characterizing the shape of the potential. $\Gamma(x)$ denotes here the complete gamma function\,\cite{Abramowitz1964}.

\begin{figure}[!t]
\centering
\includegraphics[width=8.6cm,clip=true]{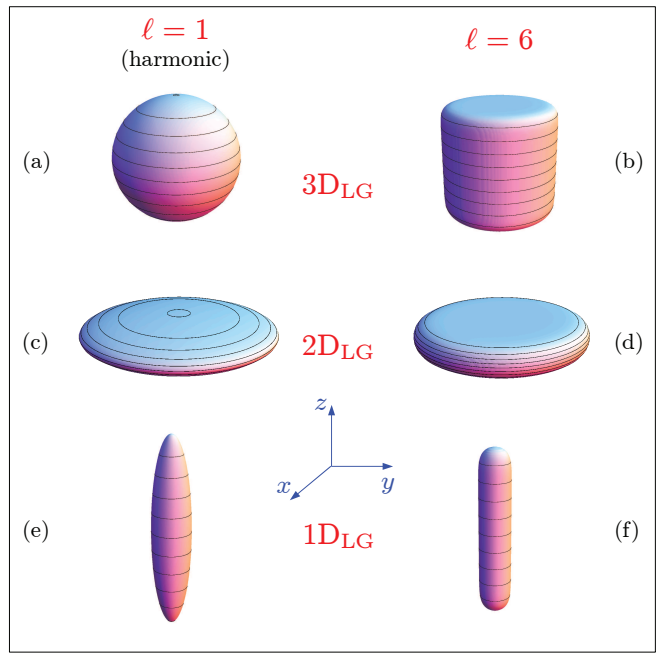}
\caption{(Color online) Plots of atomic iso-density in the 1D$_\textrm{LG}$, 2D$_\textrm{LG}$ and 3D$_\textrm{LG}$ configurations. The condensate wave function is obtained from the solution of the three-dimensional Gross-Pitaevskii equation for one million Rubidium-87 atoms in a Laguerre-Gaussian trap with $\ell=1$ (left column) and $\ell=6$ (right column). One can observe the usual cigar, disk and spherical shapes for the harmonic trap corresponding to $\ell=1$, while a kind of finger, a kind of hockey puck and a cylindrical shape are obtained in the quasi-homogeneous trap corresponding to $\ell=6$.}
\label{fig:BEC-shape}
\end{figure}

We then approximate the condensate chemical potential by its Thomas-Fermi expression (valid for large condensate occupation numbers) as
\begin{equation}
\mu_{\textrm{TF}}(N_c) = \left[ \frac{g\,\Gamma\left(\eta+\frac{3}{2}\right)}{2\pi\,C_{\alpha\beta}} \right]^{\frac{2}{2\eta+1}}\,\times\,N_c^{\frac{2}{2\eta+1}}\,.
\label{Eq:muTF}
\end{equation}
The volume $\vol_c$ of the Thomas-Fermi condensate, which corresponds to the region where $V(\rho,z) \leqslant \mu_{\textrm{TF}}(N_c)$, is then obtained as
\begin{equation}
\vol_{c} =\; \frac{2\pi\,C_{\alpha\beta}}{\Gamma\left(\eta+\frac{1}{2}\right)}\,\left[ \frac{g\,\Gamma\left(\eta+\frac{3}{2}\right)}{2\pi\,C_{\alpha\beta}}\,N_c \right]^{\frac{2\eta-1}{2\eta+1}}\,.
\label{Eq:volume-condensate}
\end{equation}
This volume $\vol_c$ will be fixed arbitrarily to $\vol_c=5.3 \times 10^{3}\,\mu$m$^3$ throughout this manuscript.

We now consider three distinct configurations that we label as 1D$_\textrm{LG}$, 2D$_\textrm{LG}$ and 3D$_\textrm{LG}$. In the 1D$_\textrm{LG}$ case, a Laguerre-Gaussian beam of index $\ell > 1$ is imposed along a single dimension, while trapping in the two others is harmonic. This corresponds to $\alpha=2$ and $\beta=2\ell$ in Eq.(\ref{Eq:Potential}). Similarly, the 2D$_\textrm{LG}$ case corresponds to a 2D power-law trapping potential, with $\alpha=2\ell$ and $\beta=2$. Finally, in the 3D$_\textrm{LG}$ configuration, the power-law potential is imposed in all directions, with $\alpha=\beta=2\ell$ and $U_{\!\bot}=U_{\!z}$. In the 1D$_\textrm{LG}$ and 2D$_\textrm{LG}$ cases, the confinement is arbitrarily supposed to be 5 times tighter in the harmonic trap compared to the power-law trap, in order to recover the usual cigar and disk shapes obtained in the case of a 1D and 2D confinement respectively.

Fig.\,\ref{fig:BEC-shape} shows the three-dimensional shapes of the condensates obtained by solving the Gross-Pitaevskii equation\,(\ref{Eq:GPE}) in these three configurations for $\ell = 1$ and for $\ell = 6$. These calculations were performed for one million Rubidium-87 atoms using the imaginary-time relaxation technique\,\cite{Kosloff1986}. Compared to the usual harmonic trap in 1D (cigar shape) and 2D (disk shape)\,\cite{Salasnich2002-Munoz2008}, one can observe that for large values of $\ell$ the entire accessible trap volume is occupied by an almost constant atomic density. Going from $\ell = 1$ to $\ell = 6$ therefore transforms the usual cigar shape in a kind of \textit{finger} and the usual disk shape in a kind of \textit{hockey puck}. In a 3D trap, this transformation results in a cylindrical shape for the condensate, while the usual spherical shape is obtained for $\ell=1$. The shape of the condensate can thus be modified between these different limits by simply tuning the azimuthal index $\ell$ of the Laguerre-Gaussian beams.

We will show hereafter that the variation of the shape of the potential has also a strong impact on the condensation temperature.

\subsection{Transition temperature}

The total number of particles $N$ can be expressed using the equation of state as\,\cite{Yan2000}
\begin{equation}
N = \frac{f}{1-f} + \frac{\vol_{T}(k_BT)}{\lambda_T^3}\;\Gamma\left(\eta+\frac{1}{2}\right)\,g_{\eta+1}(f)\,,
\label{Eq:EE}
\end{equation}
where $T$ is the absolute temperature, $f=e^{\mu/k_BT}$ the fugacity and $g_s(z)$ the Bose function of order $s$
\begin{equation}
g_s(z) = \sum_{i=1}^{\infty} \frac{z^i}{i^s}\,.
\label{Eq:Bosefunction}
\end{equation}
In the case of an ideal gas with no zero-point energy, the critical value of the chemical potential is $\mu=0$. Introducing this value in Eq.(\ref{Eq:EE}) and neglecting the condensate number $N_c = f / (1-f)$ yields the critical temperature of the ideal gas
\begin{equation}
k_BT^{0}_{c} = \left[ \frac{\sqrt{2\pi}\,\hbar^3}{m^{\frac{3}{2}}\,C_{\alpha\beta}\,\zeta(\eta+1)} \right]^{\frac{1}{\eta+1}}\,\times\,N^{\frac{1}{\eta+1}}\,.
\label{Eq:Tc0}
\end{equation}
One can already note from this equation that since $\frac{1}{2} \leqslant \eta = \frac{2}{\alpha} + \frac{1}{\beta} + \frac{1}{2} \leqslant 2$, the power governing the variation of the critical temperature with the number of atoms varies between $\frac{1}{3}$ (harmonic trap) and $\frac{2}{3}$ (homogeneous case). Note that similar results were already obtained in Ref.\,\cite{Bagnato1987} in the case of a separable trapping potential expressed in Cartesian coordinates. Controlling the Laguerre-Gauss index $\ell$ gives access to a wide range of trapping situations whose properties differ by their power-law variations in $N^{\frac{1}{\eta+1}}$.

For large condensate numbers, two-body interactions modify significantly the ideal transition temperature $T_c^0$ given in Eq.(\ref{Eq:Tc0}). The study of this effect has attracted a significant amount of attention during the last 50 years\,\cite{Andersen2004}, after it was realized that the dependence of $T_c$ on the atomic interaction strength is determined by different physical mechanisms in homogeneous and in inhomogeneous (for instance harmonic) trapping potentials. In the case of a homogeneous dilute Bose gas, it is only in 1999 that this issue has been settled\,\cite{Baym1999}.

\begin{figure}[!t]
\centering
\includegraphics[width=8.6cm,clip=true]{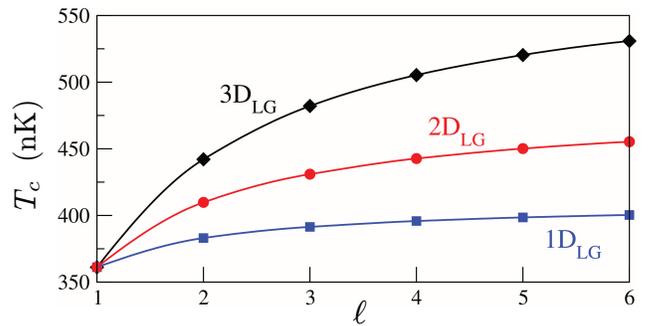}
\caption{(Color online) Condensation temperature $T_c$ [Eq.(\ref{Eq:Tc})] of one million Rubidium-87 atoms as a function of the Laguerre-Gauss index $\ell$ in the 1D$_\textrm{LG}$, 2D$_\textrm{LG}$ and 3D$_\textrm{LG}$ configurations for the same condensate volume $\vol_c = \vol_{T}(\mu_{\textrm{TF}}) = 5.3 \times 10^{3}\,\mu$m$^3$ (see text for details).}
\label{fig:temperature}
\end{figure}

For the range of parameters explored here, \textit{i.e.} in the presence of a power-law trapping potential with $1 \leqslant \ell \leqslant 6$, long-wavelength fluctuations still have a marginal impact on the critical temperature. The shift due to atomic interactions can therefore be obtained with a good accuracy using mean-field theory in the thermodynamic limit\,\cite{Zobay2004-Zobay2005}. In our particular case, the corrections due to atomic interactions can be introduced up to the second order in the parameter $q=a_s/\lambda_{T^{0}_c}$ as
\begin{equation}
T_c = \Big[ 1 + D_1 (\eta) q + D_1'(\eta) q^{2\eta} + D_2(\eta) q^2 + \ldots \Big]\,T^{0}_{c}\,,
\label{Eq:Tc}
\end{equation}
where the coefficients $D_1(\eta)$ and $D'_1(\eta)$ characterize large distance and short range potential shape effects, respectively. These coefficients, as well as the second order term $D_2(\eta)$, can be calculated as described very pedagogically in Ref.\,\cite{Zobay2004-Zobay2005}. Note that the inclusion of corrections up to second order in the interaction strength is necessary for $\eta \leqslant 1$, \textit{i.e.} when a quasi-homogeneous Bose condensate is formed\,\cite{Zobay2004-Zobay2005,Andersen2004}. Indeed, in this case, the third and forth terms in the expansion\,(\ref{Eq:Tc}) are not negligible when compared to $D_1 (\eta) q$.

Fig.\,\ref{fig:temperature} shows the variation of the interacting condensation temperature $T_c$ with the Laguerre-Gauss index $\ell$ in the 1D$_\textrm{LG}$, 2D$_\textrm{LG}$ and 3D$_\textrm{LG}$ configurations described previously for 10$^6$ atoms.

\begin{figure}[!t]
\centering
\includegraphics[width=8.6cm,clip=true]{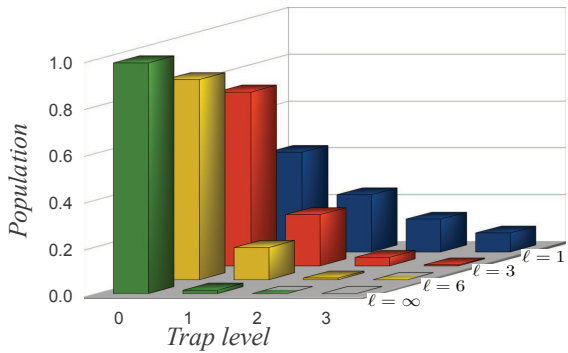}
\caption{(Color online) Bose-Einstein distribution of population in the first few 1D$_\textrm{LG}$ trap levels as a function of the Laguerre-Gauss index $\ell$. Green (front row): homogeneous case $(\ell=\infty)$. Yellow (second row): $\ell=6$. Red (third row): $\ell=3$. Blue (last row): harmonic trap $(\ell=1)$. For all values of $\ell$ the size of the ground state (\textit{i.e.} its classical turning points) is fixed at the same value, and $k_BT=2\,\hbar\omega$, where $\omega$ denotes the angular frequency of the harmonic trap.}
\label{fig:Vib-pop}
\end{figure}

One can notice in Fig.\,\ref{fig:temperature} an increase of the transition temperature with $\ell$, the effect being particularly pronounced when the power-law trap is imposed to the three dimensions. Indeed, in the 3D$_\textrm{LG}$ configuration, $T_c$ varies from 360\,nK for $\ell=1$ to 530\,nK for $\ell=6$. This large increase in $T_c$ (+50\%) is due to the different density of states $g(\varepsilon)$ obtained for $\ell > 1$ when compared to a harmonic trap. Indeed, in a semi-classical approach
\begin{equation}
g(\varepsilon) = \frac{1}{h^3} \int d^3\mathbf{r} \int d^3\mathbf{p}\; \delta\left[\varepsilon-\varepsilon_\ell(\mathbf{r},\mathbf{p})\right]\,,
\label{Eq:gE-def}
\end{equation}
where $\varepsilon_\ell(\mathbf{r},\mathbf{p})$ denotes the dispersion relation $\mathbf{p}^2/2m +V(\rho,z)$. Inserting Eq.(\ref{Eq:Potential}) in this expression and integrating over the position and momentum coordinates yields
\begin{equation}
g(\varepsilon) = \frac{1}{\hbar^3}\frac{m^\frac{3}{2}}{\sqrt{2\pi}}\frac{C_{\alpha\beta}}{\Gamma\left(\eta+1\right)}\,\varepsilon^\eta.
\label{Eq:gE}
\end{equation}
Consequently, $g(\varepsilon)$ is proportional to $\varepsilon^2$ for a harmonic trap and to $\sqrt{\varepsilon}$ in the homogeneous limit. A schematic representation of the associated energy level structures can be seen in Fig.\,\ref{fig:schematic}. A consequence of these different ladder structures is that, when the volume that the atoms can access is fixed, the population of the ground state at a given temperature is much higher for $\ell > 1$ when compared to the same population in a harmonic trap, characterized by $\ell=1$.

This can be seen in Fig.\,\ref{fig:Vib-pop} which shows, at fixed temperature, the Bose-Einstein distribution of population of the first few 1D$_{\mathrm{LG}}$ trap levels as a function of $\ell$. This simple effect of redistribution of atomic population favors the ground state as $\ell$ increases, and it therefore favors condensation at higher temperatures in Laguerre-Gaussian beams of high azimuthal orders $\ell$.

In addition, varying $\ell$ allows for the exploration of the crossover from the inhomogeneous regime ($\eta > 1$), where the interaction shift on $T_c$ is mainly due to large distance potential shape effects $(D_1 (\eta) q > D_1'(\eta) q^{2\eta})$, to the quasi-homogeneous regime ($\eta < 1$), where critical effects near the trap center dominate $(D_1 (\eta) q \leqslant D_1'(\eta) q^{2\eta})$.

\subsection{Experimental constraints}

To achieve Bose-Einstein condensation using these proposed Laguerre-Gaussian dark power-law laser traps, it is crucial to minimize photon scattering from the trapping light since this phenomenon usually constitutes a major source of heating in optical traps. To minimize the scattering rate, traps with large detunings are favorable, and dark traps have an advantage over bright traps\,\cite{Friedman2002}. In the following, we estimate the photon scattering rate in the 3D$_\textrm{LG}$ case, which provides the highest condensation temperature.

In the framework of the usual two-level approximation\,\cite{Miller1993}, the photon scattering rate can be expressed as
\begin{equation}
\eta_{\mathrm{sc}} = \frac{\langle I\, \rangle\,\Gamma_{\!\!s}^3}
                          {2\left[ I_s \Gamma_{\!\!s}^2 + \langle I\, \rangle\,\Gamma_{\!\!s}^2 + 4 I_s \delta^2 \right]}\,,
\label{Eq:etaSC}
\end{equation}
where $\langle I\, \rangle$ denotes the averaged light intensity experienced by the atoms\,\cite{Friedman2002}. We can estimate this averaged intensity from the atomic density $n(\mathbf{r})$ and the intensity profile $I(\mathbf{r})$ using the expression
\begin{equation}
\langle I\, \rangle = \frac{1}{N} \int n(\mathbf{r})\,I(\mathbf{r})\,d^3\mathbf{r}\,.
\label{Eq:Iaverage}
\end{equation}
The local density approximation\,\cite{LDA} can be used to express the atomic density of a thermal cloud of cold atoms $n(\mathbf{r})$ in Eq.\,(\ref{Eq:Iaverage}). This yields
\begin{equation}
\langle I\, \rangle = \frac{3\,2^{\ell+1} P}{\ell\,\ell!\,w_0^{2\ell+2}}\,
                      \left(\frac{kT}{U}\right)
                      \left(\frac{g_{\frac{5\ell+3}{2\ell}}\left[e^\frac{\mu}{kT}\right]}
                                 {g_{\frac{3\ell+3}{2\ell}}\left[e^\frac{\mu}{kT}\right]}\right)\,,
\end{equation}
where $U=U_{\!\bot}=U_{\!z}$, and where the chemical potential $\mu$ of the thermal cloud can be obtained by solving the integral equation
$N = \int n(\mathbf{r})\,d^3\mathbf{r}$.

\begin{figure}[!t]
\centering
\includegraphics[width=8.6cm,clip=true]{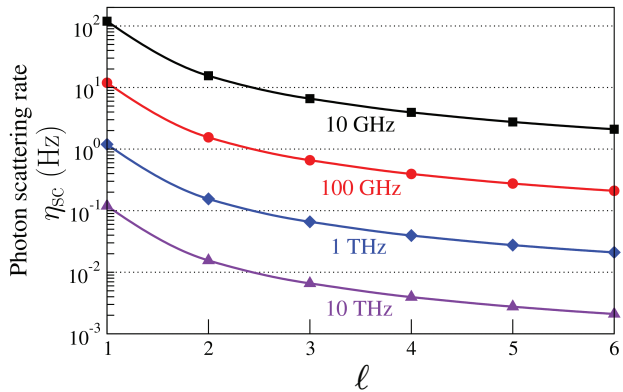}
\caption{(Color online) Photon scattering rate $\eta_{\mathrm{sc}}$ in s$^{-1}$ as a function of $\ell$ in the 3D$_\textrm{LG}$ case with $P=5$\,W for a thermal atomic cloud at $T=1\,\mu$K. Laser detuning $\delta$: black squares $2\pi \times 10$\,GHz, red circles $2\pi \times 100$\,GHz, blue diamonds $2\pi \times 1$\,THz, and violet triangles $2\pi \times 10$\,THz.}
\label{fig:scattering}
\end{figure}

The variation of the photon scattering rate with $\ell$ is shown Fig.\,\ref{fig:scattering} for a thermal atomic cloud at $T=1\,\mu$K and for different laser detunings with $P=5$\,W in the 3D$_\textrm{LG}$ case. One can see that the smallest detunings $\delta=2\pi \times 10$\,GHz and $\delta=2\pi \times 100$\,GHz yield relatively high photon scattering rates of the order of 10 to 100 events per second for $\ell = 1$. On the other hand, far detuned from the atomic resonance, with $\delta = 2\pi \times 10$\,THz for instance, the photon scattering rate in these traps is extremely low: in the range of $2 \times 10^{-3}$ to 10$^{-1}$ s$^{-1}$ only, depending on the value of $\ell$. This corresponds to heating rates in the range of 1 to 40\,nK/s. These small values make of these types of traps good candidates for evaporative cooling towards quantum degeneracy.

Moreover, traps with large values of $\ell$ are less sensitive to heating processes due to photon scattering events than harmonic traps of the same volume (corresponding to $\ell=1$ in Fig.\,\ref{fig:scattering}). This is due to a smaller value of the averaged intensity of the trapping light, as it could be expected from the qualitative picture seen Fig.\,\ref{fig:schematic}.

\begin{figure}[!b]
\centering
\includegraphics[width=8.6cm,clip=true]{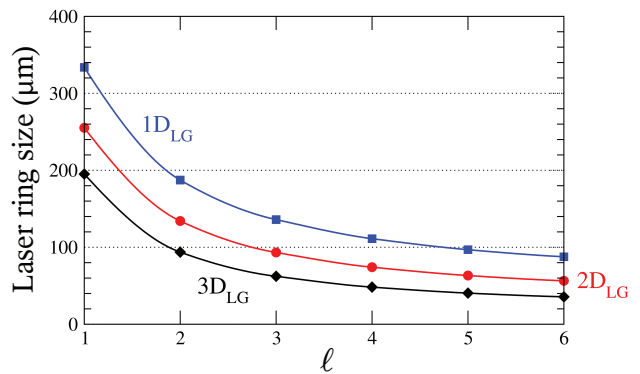}
\caption{(Color online) Laguerre-Gauss laser ring size $\rho_0=w_0\sqrt{\ell/2}$ as a function of the Laguerre-Gauss index $\ell$ for the 1D$_\textrm{LG}$, 2D$_\textrm{LG}$ and 3D$_\textrm{LG}$ configurations described in the text, with the same condensate volume $\vol_c = \vol_{T}(\mu_{\textrm{TF}}) = 5.3 \times 10^{3}\,\mu$m$^3$. The laser power is $P=5$\,W and the light detuning is $\delta=2\pi \times 10$\,THz.}
\label{fig:waist}
\end{figure}

In addition, with fixed laser parameters, increasing the Laguerre-Gauss index $\ell$ results in the formation of a less confining potential [see Eq.\,(\ref{Eq:Dipole}) or Ref.\,\cite{Mestre2010} for instance]. Maintaining a fixed condensate volume when $\ell$ is changed therefore requires to modify the laser power $P$, detuning $\delta$, or waist $w_0$. These limitations are described in Fig.\,\ref{fig:waist}. This Figure shows, for the realistic and moderate laser power $P=5$\,W and the laser detuning $\delta=2\pi \times 10$\,THz, the variation of the laser ring size $\rho_0$ of the Laguerre-Gauss beam as a function of $\ell$, in the 1D$_\textrm{LG}$, 2D$_\textrm{LG}$ and 3D$_\textrm{LG}$ configurations with $N = 10^6$ atoms.

In this Figure, one can see that achieving condensation with a fixed average atomic density requires a smaller ring size $\rho_0$ (and therefore a smaller beam waist radius $w_0$) in the case of a quasi-homogeneous trap $(\ell \gg 1)$ when compared to a harmonic trap $(\ell = 1)$.

A higher condensation temperature is therefore obtained in our study with larger values of $\ell$ at the cost of a better focusing of the beam. In all cases the ring size remains larger than 35\,$\mu$m, the beam waist radius $w_0$ always being larger than 20\,$\mu$m. The focusing required is therefore quite achievable experimentally. One can also note that a larger waist radius would be obtained if a higher laser power $P$ was chosen.

\subsection{Kinetics of condensation}

Beyond its thermodynamics properties, the kinetics of formation of the condensate may also be altered by the shape of the trapping potential.

The fundamental process which governs this kinetics is bosonic stimulation. This effect, which results from the symmetry of the bosonic wave function with respect to the interchange of any pair of particles, induces a self-acceleration of the condensate growth since the growth rate is proportional to the number of ground state atoms already present.

To determine the influence of Laguerre-Gauss optical traps on the kinetics of condensation, we adopt the quantum kinetic theory approach derived by Gardiner\,\textit{et al} in 1998\,\cite{Gardiner1998}. This approach consists in a major extension of the initial model derived in 1997\,\cite{Gardiner1997}, taking into account the evolution of the occupations of lower trap levels and of the Bose-Einstein distribution for the occupation of higher trapped levels. This model describes the growth of a condensate from a non-depletable bath of thermal atoms at a fixed positive chemical potential $\mu$ and temperature $T$. It is therefore only valid for temperatures such that at equilibrium the condensate fraction remains small $(N_c^{\textrm{eq}}/N \ll 1)$. Comparisons with more elaborate theoretical descriptions and with experimental data have shown that this approach is valid up to condensate fractions of about 10\%\,\cite{Gardiner1998,Davis2000}.

In this case one is left with a simple rate equation for the condensate number $N_c(t)$
\begin{equation}
\frac{dN_c}{dt} = 2W^{+} \times \left[ \left(1-e^{\frac{\mu_c -\mu}{k_BT}}\right) N_c + 1 \right]
\label{Eq:SGE}
\end{equation}
where the growth rate\,\cite{DavisThesis}
\begin{equation}
W^{+} = \mathscr{C}(T) \left\{ \mathscr{L}^2(\varphi) + \sum_{p=1}^{\infty} \left[ \mathscr{L}(\varphi) + \sum_{q=1}^{p} \frac{\varphi^q}{q} \right]^2 e^{p\frac{\mu_c-\mu}{k_BT}} \right\}
\label{Eq:Wplus}
\end{equation}
is an explicit function of $T$. It is also an implicit function of $N_c$ and of the trap parameters through the expression of the chemical potential of the condensate $\mu_c = \mu_{\textrm{TF}}(N_c)$ given in Eq.\,(\ref{Eq:muTF}). In the expression of the growth rate $W^+$, $\mathscr{L}(\varphi)$ is given by
\begin{equation}
\mathscr{L}(\varphi) = \ln(1-\varphi)\,,
\end{equation}
with
\begin{equation}
\varphi = \exp\left(\frac{\mu-2\mu_{\textrm{TF}}(N_c^{\textrm{eq}})}{k_BT}\right)\,,
\end{equation}
and
\begin{equation}
\mathscr{C}(T) = \frac{4m(a_{s}k_BT)^{2}}{\pi\hbar^3}\,.
\label{Eq:CT}
\end{equation}

\begin{figure}[!t]
\centering
\includegraphics[width=8.6cm,clip=true]{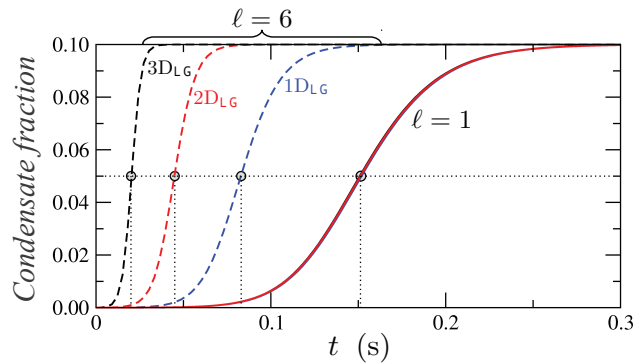}
\caption{(Color online) Condensate growth curves $N_c(t)/N$ as a function of time for an equilibrium condensed fraction $N_c^{\textrm{eq}}/N=0.1$ and $N=10^6$ in the 1D$_\textrm{LG}$, 2D$_\textrm{LG}$ and 3D$_\textrm{LG}$ configurations described in the text, with the same condensate volume $\vol_c = \vol_{T}(\mu_{\textrm{TF}}) = 5.3 \times 10^{3}\,\mu$m$^3$. The two cases considered here are the one of the harmonic trap $\ell=1$ (solid lines) and the one of the $\ell=6$ Laguerre-Gauss trap (dashed lines).}
\label{fig:dynamics}
\end{figure}

Fig.\,\ref{fig:dynamics} shows the variation of the condensed fraction $N_c(t)/N$ with time for the Laguerre-Gauss parameters \mbox{$\ell=1$} (solid lines) and 6 (dashed lines) in the 1D$_\textrm{LG}$, 2D$_\textrm{LG}$ and 3D$_\textrm{LG}$ configurations with $N_c^{\textrm{eq}}/N=0.1$ and $N=10^6$ atoms. We have assumed here that the condensate is initially unoccupied: $N_c(0)=0$.

In all cases, after a latent period during which the growth is dominated by the spontaneous rate equation $\dot{N}_c \simeq 2W^{+}$, the stimulated effect takes over, and $\dot{N}_c \varpropto 2W^{+} \times N_c$. This results in a very rapid growth, until saturation is achieved when $\mu_c \simeq \mu$.

For $\ell=1$, the 1D$_\textrm{LG}$, 2D$_\textrm{LG}$ and 3D$_\textrm{LG}$ configurations are characterized by the same shape parameter $\eta$ and by identical trap volumes. These three configurations therefore present identical density of states, and the time required to form a condensate in these three traps is the same: about 0.3\,s. On the other hand, with $\ell=6$, even if the trap volume is fixed, a shape effect appears since the shape parameter $\eta$ differs in the 1D$_\textrm{LG}$, 2D$_\textrm{LG}$ and 3D$_\textrm{LG}$ configurations. As a consequence, the times of formation of the condensate in these three traps are different. The fastest condensation takes place in the 3D$_\textrm{LG}$ configuration, in about 35\,ms. A substantial speed-up of up to one order of magnitude is therefore expected using Laguerre-Gaussian beams when compared to the usual harmonic trap. This acceleration can be interpreted using the growth equation\,(\ref{Eq:SGE}) from the increase of the condensation rate $W^+$ with temperature\,\cite{Davis2000} and from the shape effect due to the $\eta$-dependence of $W^+$.

\subsection{Evaporation}

Beyond the specific problem of the kinetics of the condensation process studied here, remains the crucial point of the mechanism used to cool the atoms in these Laguerre-Gauss power-law traps.

As mentioned previously, in a blue-detuned Laguerre-Gaussian optical trap, the trapping potential is not simply given by a pure power-law variation, but by the exact expression (\ref{Eq:Dipole-exact}). This trapping potential is therefore characterized by the presence of a potential barrier at $\rho=\rho_0$. Consequently, an evaporation ramp can be carried out to cool the atoms by simply lowering the power $P$ of the Laguerre-Gauss beam, as it is commonly done with standard Gaussian laser beams\,\cite{Adams1995}.

However, this simple evaporation mechanism suffers from the fact that the trap confinement is reduced during this forced evaporative cooling procedure. This is a serious limitation which originates from an induced reduction of the trap frequency, of the collision rate, and consequently of the evaporation efficiency. Achieving condensation using this simple forced evaporation procedure therefore requires starting from an atomic cloud with a large phase-space density and a high collision rate\,\cite{Barrett2001}.

This unwanted decrease of the evaporation rate can be limited using for instance a mobile lens in order to decrease the beam waist radius at the same time as one decreases the laser power\,\cite{Kinoshita2005}.

Recently, new strategies have also been developed in order to improve the evaporation efficiency in optical traps. One of these successful procedures uses for instance a combination of a tightly confining optical dipole trap and of a much wider laser beam in order to control independently the trap confinement and the trap depth\,\cite{Clement2009}.

There is no reason to think that achieving large initial phase-space densities could be problematic with Laguerre-Gaussian traps, but even if it is the case in a specific experimental configuration, a similar approach could be considered to improve the evaporation efficiency.

In any case, it is interesting to note that the growth rate of the condensate $W^+$ is proportional to the atomic collision rate. The increased rate of formation of the condensate with an an-harmonic Laguerre Gaussian trap LG$_{0}^{\ell}$, with $\ell > 1$, is thus primarily due to a higher collision rate in such traps at the condensation temperature. This increased collision rate should be helpful for BEC production.

\section{Conclusion}

To conclude, we have presented a realistic theoretical analysis of an original all-optical setup designed for Bose-Einstein condensation. This setup is based on crossed Laguerre-Gaussian laser beams. Our analysis is made in realistic experimental conditions: the results presented here require for instance 5\,W\,@\,760.4\,nm, with a laser detuning $\delta=2\pi \times 10$\,THz and with laser ring sizes $\rho_0 \geqslant 35\,\mu$m, corresponding to laser waist radius $w_0 \geqslant 20\,\mu$m. In these trapping potentials, the photon scattering rate can be reduced down to extremely low values, of the order of a few 10$^{-3}$\,s$^{-1}$, thus providing long coherence times for the trapped atoms.

We have shown that high Laguerre-Gauss azimuthal orders $\ell$ provide increased condensation temperatures (+50\%) when compared to the usual harmonic trapping situation. In this case, the condensate formed for different values of $\ell$ has the same typical size but has a different shape. Furthermore, a substantial speed-up (up to one order of magnitude) for the time of formation of the condensate is also predicted. These improvements, whose physical origin lies in the density of states associated with these traps, along with the long coherence times expected in dark optical traps, could be influential in domains where large condensate occupation numbers are necessary, or where higher experimental repetition rates are desired.

In addition, this all-optical trapping configuration should allow for the experimental exploration of the crossover between a quasi-homogeneous Bose gas and an inhomogeneous one formed for instance in a harmonic trap. For $\ell > 10$, one is left with an almost perfect homogeneous Bose gas, and the influence of long-wavelength non-perturbative critical fluctuations on $T_c$ should become measurable.

Finally, compared to recent experimental realizations of Bose-Einstein condensates in a box realized with Hermite-Gaussian TEM$_{01}$ laser beams\,\cite{Meyrath2005}, our proposal should lead to a much greater steepness of the repulsive walls, and therefore to a better experimental modeling of a homogeneous trap.

\begin{acknowledgments}
The authors would like to acknowledge financial support from ANR (Agence Nationale de la Recherche, Project Number ANR-07-BLAN-0162-02), from the R\'egion Ile-de-France (Programme R\'egional SETCI), from IFRAF (Institut Francilien de Recherche sur les Atomes Froids) and from LUMAT (F\'ed\'eration LUMi\`ere-MATi\`ere du CNRS). EC acknowledges stimulating discussions with Matthew Davis, from the University of Queensland, Australia.
\end{acknowledgments}



\begin{thebibliography}{99}
\bibitem{NoteLG}
The number of radial nodes in the mode profile is \mbox{$(p+1)$} and $\ell$ is the number of $2\pi$ phase cycles around the circumference of the mode.
\end{thebibliography}
\end{document}